\documentclass[12pt]{article}
\usepackage{graphicx}
\usepackage{amssymb}

\newcommand{\beq}{\begin{equation}}

\newcommand{\eeq}{\end{equation}}
\newcommand{\bea}{\begin{eqnarray}}
\newcommand{\eea}{\end{eqnarray}}

\newcommand{\La}{\Lambda}

\renewcommand{\a}{\alpha}

\newcommand{\Del}{\Delta}

\renewcommand{\k}{\kappa}

\begin{document}

\begin{center}
\vspace{24pt}
{ \Large \bf Geometry of the quantum universe}

\vspace{30pt}

{\sl J. Ambj\o rn}$\,^{a,c,d}$,
{\sl A. G\"{o}rlich}$\,^{b}$,
{\sl J. Jurkiewicz}$\,^{b}$,
and {\sl R. Loll}$\,^{c,d}$

\vspace{24pt}
{\footnotesize

$^a$~The Niels Bohr Institute, Copenhagen University\\
Blegdamsvej 17, DK-2100 Copenhagen \O , Denmark.\\
{ email: ambjorn@nbi.dk}\\

\vspace{10pt}

$^b$~Institute of Physics, Jagellonian University,\\
Reymonta 4, PL 30-059 Krakow, Poland.\\
{ email: jurkiewi@thrisc.if.uj.edu.pl,
atg@th.if.uj.edu.pl}\\

\vspace{10pt}

$^c$~Institute for Theoretical Physics, Utrecht University, \\
Leuvenlaan 4, NL-3584 CE Utrecht, The Netherlands.\\ 
{ email:  r.loll@uu.nl}\\

\vspace{10pt}

$^d$~Perimeter Institute for Theoretical Physics, \\
31 Caroline St. N., Waterloo Ontario, Canada N2L 2Y5.\\ 

\vspace{10pt}

}
\vspace{48pt}

\end{center}


\begin{center}
{\bf Abstract}
\end{center}
A quantum universe with the global shape of a (Euclidean) de Sitter spacetime 
appears as dynamically generated
background geometry in the causal dynamical triangulation (CDT) 
regularization of quantum gravity. 
We investigate the micro- and macro-geometry of this universe, using
geodesic shell decompositions of spacetime. More specifically, we focus 
on evidence of fractality and global anisotropy, and on how they depend
on the bare coupling constants of the theory.

\vspace{12pt}
\noindent


\newpage

\section{Introduction}\label{intro}

The attempt to quantize gravity using conventional 
quantum field theory has been gaining considerable momentum due to 
the progress in using renormalization group techniques \cite{RG},
the understanding that one may consider an enlarged class of theories
like the ``Lifshitz gravity" suggested by P.\ Ho\v rava \cite{horava1},
and by the success of nonperturbative lattice gravity theory in terms of
Causal Dynamical Triangulations (CDT)
in reproducing some of the infrared features of our universe from 
first principles \cite{emerge,blp,semi,agjl,bigs4} (see also \cite{recent} for
recent reviews and \cite{contemp} for a non-technical account).

The lattice approach has the potential to provide a foundation 
for all other, continuum-based approaches, in the same way as lattice field
theory serves as an underlying non-perturbative definition of continuum 
quantum field theories, as emphasized by K.\ Wilson.
The lattice formulation is based on piecewise
linear geometries, which require no coordinate systems: 
all geometric information is contained in tables listing elements in
the immediate neighbourhood of a given (sub-)simplex and 
length/volume assignments for specific (sub-)simplices. 
This provides an explicit realization of the relational nature of aspects of
pure spacetime geometry, familiar from the classical
theory of General Relativity.

While getting rid of coordinates and the redundancy associated with
the arbitrariness of a coordinate choice is highly attractive from a conceptual
point of view, it comes with a unique set of challenges when one tries to
extract information about the quantum geometry. 
The well-known problem of identifying invariantly defined gravitational 
observables is aggravated in the quantum theory,
where in addition one has to consider expectation values, that is,
averages over all geometries.
If instead one wanted to work with individual path integral configurations,
which do have a definite geometry,
great care has to be taken in interpreting them since they are not
physical, in the same way as an individual path in the path integral 
of the particle is not an observable. Moreover, the
typical configurations in a quantum field theory which is not topological
are dominated by ultraviolet fluctuations, making it rather subtle to
extract physical information from them.
     
In this letter we will define and measure a number of ``quantum observables" 
to quantify further the geometric properties of the quantum de Sitter universe to
emerge from nonperturbative lattice simulations of quantum gravity in
terms of causal dynamical triangulations. Building on previous results
presented in \cite{blp,agjl,bigs4}, the quantities we will consider
characterize the fractality of the spacetime as a whole and of certain hypersurfaces
inside it, as well as potential global anisotropies between the time and space
directions of the quantum universe.

\section{The macroscopic $S^4$-universe}

In practice, the nonperturbative and background-independent 
quantization of gravity in CDT proceeds via
Monte Carlo simulations, which generate a sequence of piecewise flat spacetime 
geometries from the following input: (i) a choice of global spacetime topology,
usually $S^3\times S^1$, (ii) the Einstein-Hilbert
action, which has a natural implementation on piecewise linear geometries,
(iii) a specific form of the piecewise linear, Minkowskian building blocks, 
the four-simplices to be glued together, (iv) a global proper-time 
foliation for each path integral history, and, finally, 
(v) a rotation to Euclidean time of each path integral configuration
(see \cite{blp} for technical descriptions of all of the above). 
On the output side, we observe the emergence of a background geometry,
with well-defined quantum fluctuations around it, whose large-scale shape
has been matched with great accuracy to that of a ``round $S^4$", a (Euclidean)
de Sitter universe, see \cite{bigs4} for details.

This is a result which is (a) non-trivial, and (b) not universally true. It
is non-trivial because the four-sphere is only a saddle point solution to the Euclidean
equations and there is no obvious reason why it should dominate the path integral,
in particular, since the action is unbounded from below\footnote{More precisely,
the bare action is bounded below due to the lattice regularization, but in
taking the continuum limit it can become arbitrarily large and negative.}.
This means that the appearance of $S^4$ is due to a subtle interplay between 
the entropy of configurations (the path integral measure) and the bare 
action. This is also the reason why the result is not universally true: only in a 
certain range of bare coupling constants will the $S^4$-like background
dominate. It is the geometries in this so-called ``phase C" \cite{blp} whose
properties we will investigate presently.
For other values of the bare coupling constants 
one finds other phases (called $A$ and $B$ in \cite{blp}), 
and phase transitions between them. We not in passing that
the phase diagram of CDT quantum
gravity bears an intriguing resemblance to that of Lifshitz gravity, as discussed
in some detail in \cite{to-app}, and is the subject of ongoing
research. 

The Euclidean Einstein action and its implementation on piecewise linear 
geometries are given by
\bea
S_E&=& \frac{1}{16\pi^2 G} \int \sqrt{g} (-R+2\La) \nonumber \\
 &\to & -(\kappa_0+6\Delta) N_0+\kappa_4 (N_{4}^{(4,1)}+N_{4}^{(3,2)})+
\Delta (2 N_{4}^{(4,1)}+N_{4}^{(3,2)}),
\label{actshort}
\eea 
where $N_0$ is the number of vertices, $N_4^{(4,1)}$ 
the number of four-simplices with four vertices in one spatial slice and 
one vertex in one of the adjacent spatial slices, and $N_{4}^{(3,2)}$ 
the number of four-simplices with three vertices in one spatial slice 
and two vertices in
a neighbouring slice. The coupling $\k_0$ is proportional to the inverse bare
gravitational coupling constant, while $\k_4$ is linearly related to the bare
cosmological coupling constant. Finally, $\Del$ is an asymmetry parameter
related to the fact that we allow for a finite relative scaling between the length
of space- and time-like links. The reason why the simplicial form of the 
action is so simple is that all building blocks of type (4,1) and all
building blocks of type (3,2) are identical.

Since for simulation-technical reasons it is preferable to keep the total 
four-volume fixed during the Monte Carlo simulation, $\k_4$ effectively
does not appear as a coupling constant. Instead, one can perform
simulations for different four-volumes. This leaves us with two bare 
coupling constants, $\k_0$ and $\Del$. We start out with 
$(\k_0,\Del) = (2.2,0.6)$, 
a value firmly placed in phase $C$, at which
most of our previous computer simulations were performed.
For this ``generic" value
we have shown that by a suitable global rescaling of the continuum 
proper time, the extended quantum universe can be viewed as a standard
round four-sphere with superimposed quantum fluctuations \cite{agjl,bigs4}. 
We also reported in \cite{bigs4} that the overall shape of the background 
universe can change as a function of the bare coupling constants. In other
words, to map the dynamically generated quantum universe 
at generic points in phase $C$ to a $S^4$ requires a coupling 
constant-dependent rescaling of ``time" with respect to ``space". 

\begin{figure}
\centering
\includegraphics[width=0.75\textwidth]{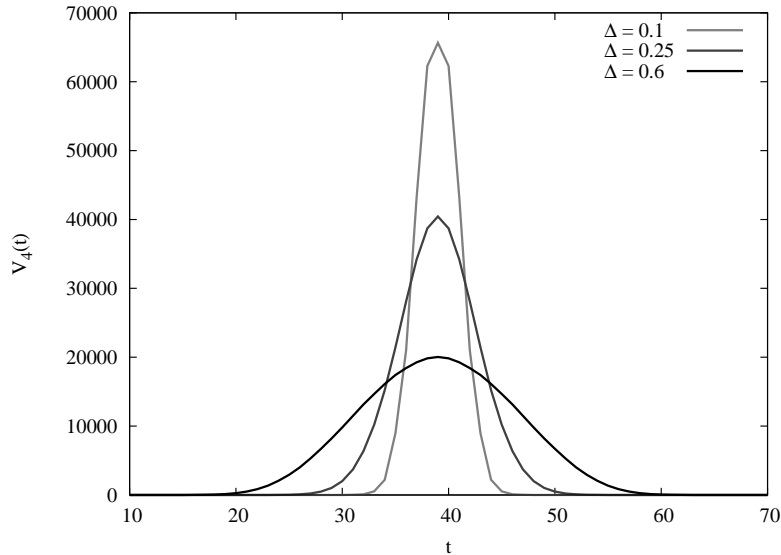}
\caption{\label{fig1} The volume profile $\langle V_4(t)\rangle$ 
for decreasing values of the asymmetry $\Del$ = 0.6, 0.25 and 0.1,
with total four-volume kept constant. The profile narrows as
$\Delta$ decreases.}
\end{figure}

In this article we are mainly interested in the 
changes that occur when one decreases $\Del$. 
The reason is that in this manner one approaches the 
phase transition between phases C and B, which is a 
potential candidate for a second-order transition line.\footnote{By contrast,
the transition 
between phases C and A, reached by keeping $\Del$ fixed and changing 
$\kappa_0$ is firmly first order and therefore less interesting from the 
point of view of obtaining new continuum physics, see \cite{to-app} for a
more detailed discussion.} 
To give a first indication of what happens, we have measured the
change in shape, by which we shall mean the average volume
profile $\langle V_4(t)\rangle$ as a function of the lattice proper time $t$,
where $V_4(t)$ denotes the (discrete) four-volume ($=$number of
four-simplices) located in the 
spacetime slab between times $t$ and $t+1$.\footnote{In the range
of coupling constants considered here, $V_4(t)$ is essentially proportional to 
the three-volume distribution $V_3(t)$, defined as the number of spatial 
tetrahedra at time $t$ (those which lie entirely inside spatial hypersurfaces of
constant $t$), whose distribution was discussed previously in \cite{agjl,bigs4}.}
Fig. \ref{fig1} illustrates how the universe's extension in the time direction,
measured in units of discrete lattice steps,
becomes shorter when we decreas $\Del$ from 0.6 to 0.1.
Our main aim in the present work is to describe
the geometry of CDT's quantum universe in greater quantitative detail, both 
at the generic point and when changing $\Del$.

\section{Exploring the universe by shell decomposition}\label{three}

To study the invariant properties of geometry we move along geodesics,
which in the Euclideanized, piecewise linear context we {\it define}  
as the shortest piecewise straight paths between any two centres of 
four-simplices, 
where each path consists of a sequence of straight segments connecting the 
centres of neighbouring four-simplices. The length of a path is simply
taken as the number of ``hops" between adjacent four-simplices. 
We could in principle work with a
finer-grained definition, for example, by using some notion of geodesic
on the piecewise linear geometry, which is defined between any pair of 
points.
However, there is no reason for the details of a piecewise linear geometry
of an individual triangulation to be of interest at the very shortest 
distances, where they are clearly an artifact of the specific choice of building 
blocks made. We expect that our simple definition of a geodesic will in the
continuum limit (that is, on scales sufficiently large compared to the cut-off
scale) lead to the same geometric results as other ``reasonable" 
definitions.\footnote{An explicit example of this mechanism has
been analyzed in two dimensions,
where the use of link distance between pairs of
vertices of a triangulation can be compared with that of
triangle distance between centres of triangles, analogous to what
we are using here in four dimensions. The latter was shown to be on
average several times larger, but universal distributions characterizing
the quantum geometry in the continuum were identical \cite{distances}.}
Note that our way of implementing geodesics does not distinguish between
segments in space and time direction; we will return in
later sections to the issue of relative scaling
between spatial and time distances. 

The way we will make use of geodesics in the present article is by
propagating either from a point or a given hypersurface in discrete 
geodesic steps of unit length to consecutive geodesic shells, foliating
part or all of the universe. We collect certain data associated with such
shell decompositions, from which we reconstruct the following geometric 
information about the quantum universe: (i) its fractal structure,
(ii) its average volume distribution as function of time and 
spatial distance, and (iii) an estimate of its global shape.
While (ii) and (iii) refer to genuine averages over many configurations, 
measurements relating to (i) will make use of individual configurations in the way 
described below.

Let us describe some key elements of our measurement process. 
For any fixed, given universe configuration generated by the Monte Carlo
simulation, we first locate its (non-unique) ``centre", defined as any four-simplex
lying in the spacetime slab with maximal volume $V_4(t)$, whose time label
we will denote by $t_0$. Picking an arbitrary four-simplex in this maximal
slab, we move outwards from this centre in 
spherical shells (in a four-dimensional
sense), advancing in geodesic steps of length 1.  
We record various pieces of information on the way, most prominently, 
the four-volume $V_4(t,r)$ in the shell of four-simplices located 
a distance of $r$ steps away from the centre and at the same time 
located in time
slab $t$. Thus $V_4(t,r)$ constitutes a fraction of both $V_4(t)$ and
of $\tilde V_4(r)$, which by definition is the four-volume of the entire shell
(i.e. the number of four-simplices contained in it) at distance $r$. 
Summarizing the situation, we have the relations
\beq\label{3.1}
V_4(t) = \sum_r V_4(t,r)
\eeq
and 
\beq\label{3.2}
\tilde V_4(r) = \sum_t V_4(t,r),
\eeq
as well as
\beq
V_4=\sum_t V_4(t)\equiv \sum_r \tilde V_4(r)
\eeq
for the total four-volume $V_4$.
In order to be able to average over many configurations we redefine
our time labeling such that $t_0$ always corresponds to time zero. 
For investigating the fractal nature of an individual spacetime configuration
(its ``branchedness" and possible associated self-similarity), 
we also record the {\it connectivity} of the shell at distance $r$, where 
two simplices in the shell at distance $r$  
are called connected if one can find a path connecting them using simplices only 
from shells with some larger distance $r' \geq r$.\footnote{Note that this
definition is suitable for a simply connected space, like the effective $S^4$ 
we are dealing with here (we never consider geodesics that wind around
the time direction in a noncontractible manner).}

\section{The results}

\subsection{The fractal structure}

\begin{figure}[t]
\centering
\includegraphics[height=9cm]{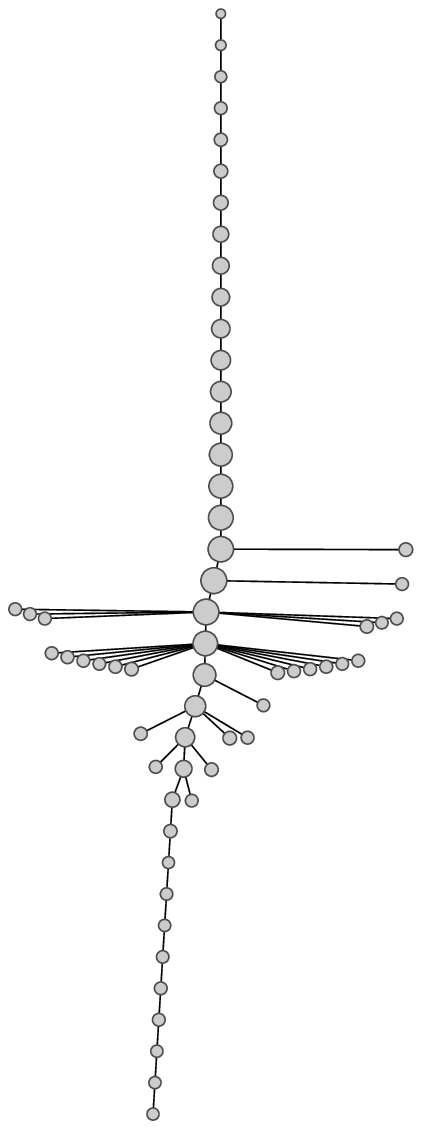}
\includegraphics[height=9cm]{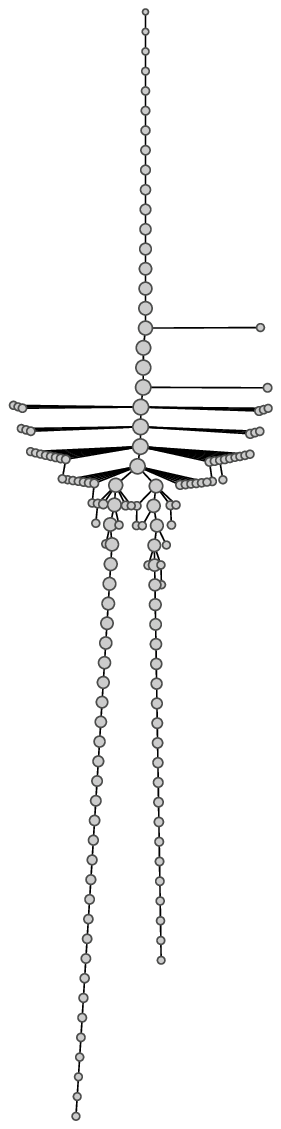}
\includegraphics[height=9cm]{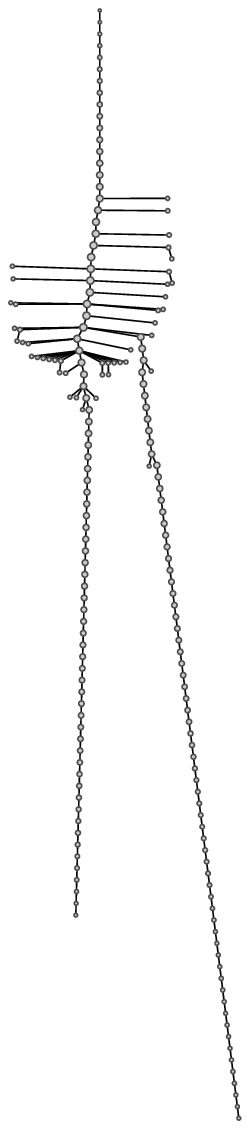}
\caption{\label{fig2} Tree graphs illustrating the connectedness of radial
shells, starting from a central four-simplex of a spacetime configuration
(top node of graph) and moving outward in discrete, concentric 
spherical shells (corresponding to going down or sideways in the tree graph).
From left to right, measurements taken at $\Del$= 0.06, 0.25 and 0.6.}
\end{figure}

We have measured the fractal structure of individual path-integral 
histories for a number
of different values of $\Del$, starting at $\Del = 0.6$ and ending
at $\Del =0.06$ (with $\kappa_0=2.2$ understood), with 
Fig.\ \ref{fig2} giving a representative sample of data. 
The figures should be read as follows: the top node of each of the three
tree graphs represents the chosen ``centre of the universe", and distance 
from the centre increases as one moves down or sideways in the graph. 
Each node represents a connected component
of a shell. A line connecting two nodes indicates that there is
a four-simplex in one of the connected components which is a direct
neighbour of a 
four-simplex in a connected component in a neighbouring shell. By construction, 
such a graph will be a tree graph. One might call it a diffusion tree, since one 
essentially follows the front of the simplest diffusion process
one can study on a given configuration. The size of a node is a measure of
the number of four-simplices in the associated connected component of the shell.
To optimize the graphical presentation, the radius $\rho$ of the node has been
chosen as $\rho \propto V_4^{1/10}$, and only shells 
(and sub-shells) with more than 40 simplices have been included.   

The qualitative features of the graphs shown in Fig.\ \ref{fig2} are generic 
and there is little sign of any nontrivial branching structure.
One typically finds one sequence of connected shells which dominates
(some small disconnected components are created but end almost
immediately), which at some point {\it bifurcates} into two. The figure
also illustrates that this bifurcation becomes more pronounced for larger 
$\Del$. The interpretation of this phenomenon in terms of the overall shape
of the quantum universe should be clear: for large $\Del$ we have 
a four-sphere which has been stretched along the time direction. 
The analogue in two dimensions lower would be that of a round two-sphere
stretched along one of its directions to create a prolate spheroid 
(the surface of revolution obtained by rotating an ellipse about its major
axis). Starting a diffusion process from any point along the equator
along the nonstretched direction, the diffusion front will propagate in
concentric circles until it meets itself at the antipode of the starting point,
where it will then bifurcate and move in 
opposite directions toward the pointed ends of the spheroid. 
Returning to the case of four dimensions, 
the spheroid becomes more spherical with decreasing $\Delta$, 
and consequently the bifurcation becomes less pronounced
(and even disappears for the lowest value of $\Del$), as we have been
observing. Although our discussion of fractal behaviour above is 
of a qualitative
nature, it is reasonably straightforward to construct related quantum
observables whose expectation value on the ensemble of all spacetimes
is well defined and can be measured quantitatively. An example of this is
the ``sphericity" of the universe, which will 
be introduced in Sec.\ \ref{spheri} below.

\subsection{The volume of shells}\label{volshell}

The picture of a diffusion process on a spheroid is corroborated by 
measurements of $V_4(t,r)$ for various values of $\Del$. 
Unlike in the previous section, the data collected do not refer to a 
single configuration, but to a combined average over configurations
and initial points (where for each given spacetime configuration, we
selected 100 different starting points in the maximal time slab and
repeated the diffusion process).   

\begin{figure}
\centerline{
\scalebox{0.37}{\rotatebox{0}{\includegraphics{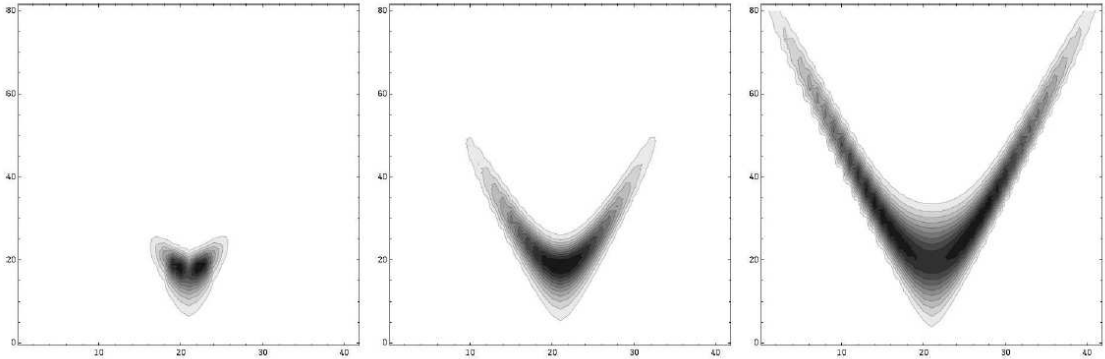}}}}
\caption{\label{fig3} Contour plots of the distribution $V_4(t,r)$, as function
of the spacetime slab $t$ (horizontal axes) and the distance $r$ travelled
by the diffusion front (vertical axes) from its initial point. 
The processes are centred in slice $t_0=20$, and the plots are taken
at $\Del$ = 0.06, 0.25 and 0.6 (left to right).}
\end{figure}

In Fig.\ \ref{fig3} we show contour plots of the distribution 
$V_4(t,r)$ for various $\Del$, which for large $\Del$ assume a
characteristic ``V''-shape in the $t$-$r$-plane. They indicate 
that the diffusion front
splits into two after a certain distance $r_{bif}$, called the {\it bifurcation 
distance}, which should be identified with half the length
of the equator of the (unstretched, round) three-sphere at time $t_0$. 
As $\Del$ decreases the V-shape diminishes, with the obvious interpretation
that the shape of the spheroid becomes more and more spherical, with
approximately equal extension in spatial and time directions.

Additional evidence for this geometric interpretation comes from 
starting the diffusion instead at one of the tips of the spheroid. 
In this case one never observes
any V-shape, and the front of diffusion is not very different from the 
proper-time slicing which was used to define the original global time 
in the computer simulations.   

A related study has recently been performed in three-dimensional 
CDT\footnote{For a definition of Causal Dynamical Triangulations 
in three dimensions, see \cite{3dcdt,d4}.}, 
using the full diffusion equation \cite{bh} and comparing it 
with the diffusion on an elongated sphere. The conclusion was that
from the point of view of long-distance diffusion one can 
indeed view the quantum geometry in the three-dimensional case as 
that of a stretched
sphere with small superimposed quantum fluctuations. In the 
four-dimensional case we have one more coupling constant at our disposal,
the asymmetry factor $\Del$, which seemingly allows us to monitor
the shape of the universe when described in terms of lattice spacings.
However, the conclusion that the real quantum universe changes shape
under a change in $\Delta$ may be premature:
as we will explain further in Sec.\ \ref{discuss}, when discussing the situation
in terms of actual physical distances (instead of just lattice units), the shape 
of the universe may actually change very little or even not at all.

\subsection{The function $\tilde V_4(r)$ and sphericity}\label{spheri}

\begin{figure}
\centering
\includegraphics[width=0.75\textwidth]{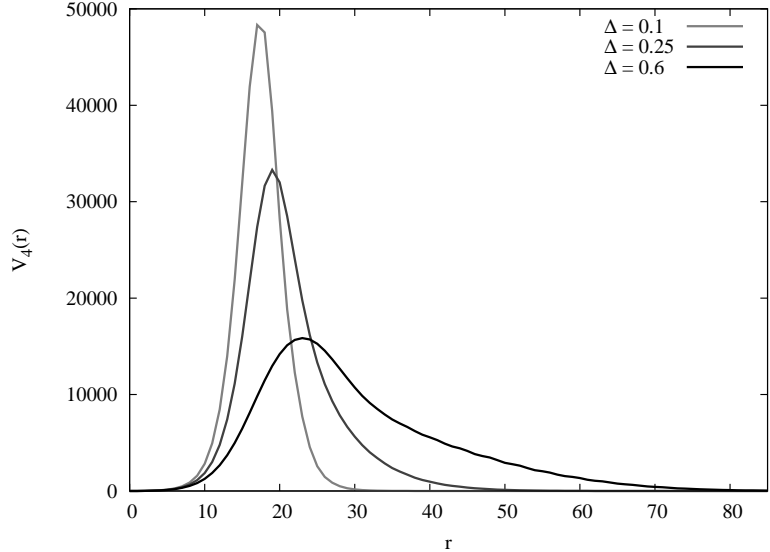}
\caption{\label{fig4} Average radial volume distribution $\tilde V_4(r)$ 
as function of the distance $r$ from a ``centre of the universe".}
\end{figure}
Next we turn to the distribution of the number $\tilde V_4(r)$ of 
four-simplices in a shell at distance $r$ from a given centre of 
the universe, as defined in Sec.\ \ref{three} above. Fig.\ \ref{fig4} 
shows $\tilde V_4(r)$ for various values of $\Del$.
For the smallest values of $\Del$ the peak is nicely symmetric
and well approximated by $A\sin^3(r/B)$, in agreement with 
earlier studies of the curve $V_4(t)$. The hypersurfaces of
constant radius $r$ are
of course completely different from those of constant $t$, but 
nevertheless it turns out that $\tilde V_4(r)$ agrees with $V_4(t)$ 
up to a rescaling of the constants. 
For larger values of $\Del$ the situation is different in that
$\tilde V_4(r)$ has a large-$r$ tail, 
which cannot be fitted to $A\sin^3(r/B)$. 
This behaviour is again consistent with universes of the form of
prolate spheroids: for a genuinely spherical configuration,
$\tilde V_4(r)$ would vanish for radii $r$ larger than the distance 
between antipodal
points. By contrast, for elongated configurations the diffusion front
bifurcates when the antipodal point is reached, 
and continues further towards the tips of the spheroid.
  
\begin{figure}
\centering
\includegraphics[width=0.75\textwidth]{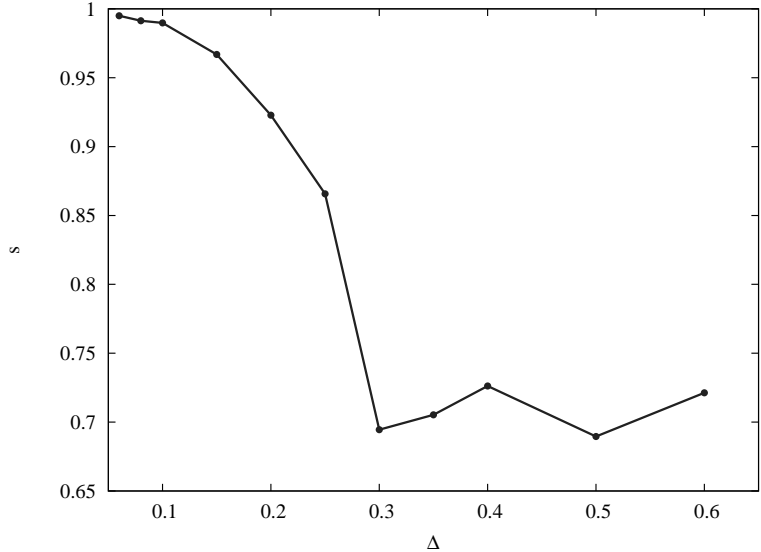}
\caption{\label{fig5} Sphericity $s$ of the quantum universe for 
different values of the asymmetry parameter $\Delta$.}
\end{figure}

Let us try to quantify how spherical our averaged configuration is by 
defining the {\it sphericity} $s$ by
\beq\label{3.3}
s := \frac{\sum_{r=0}^{r_{bif}}\tilde V_4(r)}{\sum_{r=0}^{r_{max}}\tilde V_4(r)}.
\eeq
In agreement with our earlier characterization,
$r_{bif}$ is defined operationally as the largest radius $r$ for which 
$V_4(t_0,r)$ is larger than some cut-off, here taken to be 4 (recall
that $t_0$ marks the spacetime slice of maximal four-volume).  
Similarly, $r_{max}$ is the largest distance $r$ for which $\tilde V_4(r)$ is 
larger than the cut-off. This implies that
the denominator of the quotient (\ref{3.3}) is an overall normalization,
given by the total
four-volume minus the volume of the ``stalk" (where the spatial
universe only persists because we do not allow it to shrink to zero volume).
Fig.\ \ref{fig5} shows how $s$, averaged over both configurations and
initial points, depends on $\Del$. 
From comparing with the diffusion trees one would expect $s$ to be
close to 1 for the smallest values of $\Del$ considered here. 
This is indeed what we find confirmed here. Of course, $s=1$ is exactly
the value which we would also obtain for a round sphere in the continuum.

\section{The fractal structure of spatial slices}

When studying the connected components of a shell at distance $r$
above, the connectivity was defined in a four-dimensional sense, in that
the connecting paths were allowed to lie not only in the
shell $r$, but also in shells with $r'>r$. The resulting tree structure did
not exhibit any fractality. This picture changes drastically when we 
confine ourselves to a shell at fixed $r$ and define connectivity
with respect to paths that lie entirely within that shell. 
What we have found is that the structure of the shells at fixed radius $r$ is
quantitatively similar to both that of the four-dimensional
time slabs labeled by time $t$, as well as that of the three-dimensional 
hypersurfaces at constant time $t$, made exclusively from spatial tetrahedra. 
Since the spatial hypersurfaces at fixed time $t$ are the easiest to handle 
numerically, we have used them to collect our data set. What we would like
to emphasize is that the structure reported below is equally valid for
any of the hypersurfaces or slabs appearing above, and presumably
reflects the properties of generic (reasonably chosen) hypersurfaces. 

A hypersurface of this kind is a three-dimensional triangulation, more precisely,
a piecewise linear manifold of topology $S^3$. 
The mismatch between the measured values of its spectral dimension,
$d_S \approx 1.5$, and its Hausdorff dimension, $d_H \approx 3$,
is an indicator of the nonclassical, fractal nature of these slices  
(for definitions and results, see \cite{blp}).
We will now quantify their fractality in a more direct way, with the
tree structure defined in terms of so-called ``minimal necks" 
\cite{minbu1,minbu2}. Such a {\it minimal neck} consists of four
neighbouring triangles which are glued together in such a way as to
form a minimal representation
of a topological two-sphere, or, equivalently, the surface of a solid tetrahedron,
but {\it without} the interior of the tetrahedron forming part of the 
three-dimensional triangulation. 

\begin{figure}[t]
\centering
\includegraphics[height=0.6\textwidth]{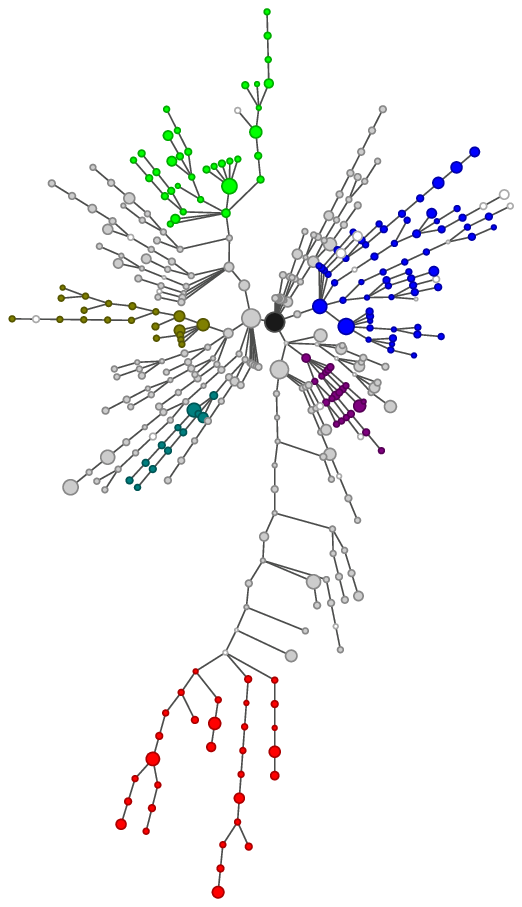}
\includegraphics[height=0.66\textwidth]{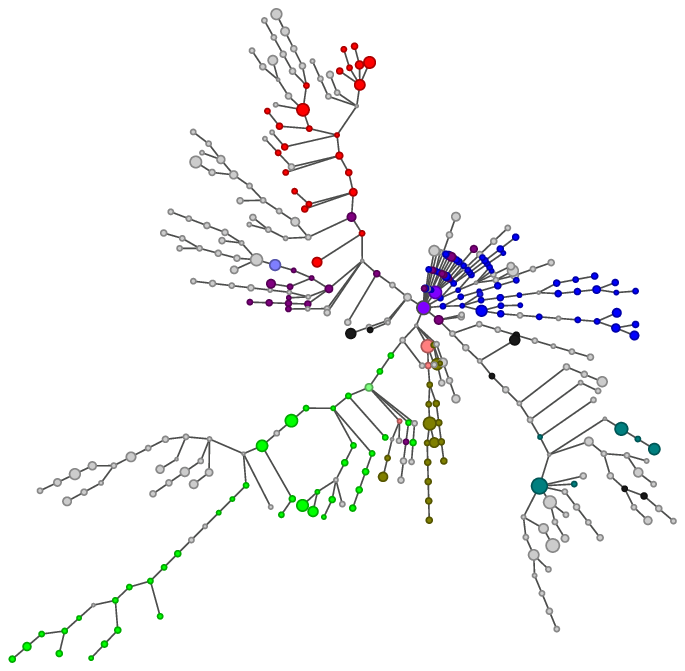}
\caption{\label{fig6} The fractal structure of two neighbouring hypersurfaces
at times $t$ and $t+1$, 
using the tree structure induced by minimal necks. Matching colours signify 
neighbouring vertices in the adjacent hypersurfaces, as explained in the text.}
\end{figure}

Cutting a triangulation along a minimal neck
will separate it into two disconnected parts which can both be
made into triangulations of $S^3$ by closing off the two boundaries, 
each given by a copy 
of the minimal neck, with two tetrahedral building blocks.\footnote{The analogous
process in two instead of three dimensions, where the minimal neck consists
of three edges forming the boundary of a triangle, is somewhat easier to
visualize.} Cutting along each minimal neck 
in the triangulation and repeating the process leaves us with a number of 
$S^3$-components. Each of these
we represent by a graph vertex, which is then reconnected by a graph edge
to each vertex representing a $S^3$-component that was originally connected
to the first one by a minimal neck. By this ``minimal-neck surgery" 
we can associate a unique tree graph to any three-dimensional triangulation. 
Fig.\ \ref{fig6} illustrates a typical tree structure associated with a given
triangulated hypersurface.
 
The resulting tree structure reflects a rough 
three-dimensional distance hierarchy, but does this persist in a
four-dimensional sense once we re-allow for paths which
can leave the hypersurface, and may lead to short-cuts between points
on it? To some extent it does, at least in a statistical sense. 
Namely, we have checked that typical
distances between pairs of points are not drastically altered when 
considering the full, four-dimensional embedding. 
The observed fractal structure is therefore not entirely an 
artifact of defining a hypersurface in a generic spacetime configuration 
appearing in the path integral, 
which of course is subject to wild ultraviolet fluctuations.

There is another type of measurement we have made to corroborate
the persistence of the fractal structure when going from one hypersurface to
the next. In order to compare the tree structures at times $t$ and $t+1$,
we define a neighbourhood relation as follows.
For a given tetrahedron belonging to a vertex 
in the constant-time slice at $t$, its ``neighbours'' in time
slice $t+1$ are the tetrahedra which are at a (four-dimensional) distance 
4 away from it\footnote{We use the distance as defined at the beginning 
of Sec.\ \ref{three} and associate to each spatial tetrahedron the unique
four-simplex it belongs to in the forward time direction, say.
The shortest possible distance 
between two time-slices is 4, since we have to go from a (4,1)- to a (3,2)- to 
a (2,3)- to a (1,4)-simplex to cross a time slab.}. 
Two $S^3$-components represented by vertices $v$ and $v'$
at times $t$ and $t+1$ are neighbours if the minimal
distance between pairs of tetrahedra $(\tau,\tau')$, $\tau\in v$, $\tau'\in v'$,
is 4. One can now study
whether this nearest-neighbour association between components (vertices) 
has any relation
with the tree structure defined {\it a priori} at time $t+1$. 
For purposes of illustration we have marked all vertices
of a given tree branch at time $t$ in a chosen colour. For each 
vertex we then find its neighbouring vertices in time slice $t+1$, and 
mark them with the same colour, allowing also for mixing of colours. 

An example of the coloured connectivity trees of two adjacent slices
is given in Fig.\ \ref{fig6}, where one should keep in mind that the
only relevant feature conveyed is the connectivity along the edges of the
graph, and not the location of the vertices in the two-dimensional plane. 
The resemblance of the two coloured graphs shows clearly that the associated 
fractal structures of the hypersurfaces are correlated, and not totally independent
of each other. 
In principle one can try to establish similar relations between slices that are
two or more steps apart, by using ``neighbours of neighbours", but this
notion of neighbourhood soon becomes rather weak. -- 
As usual in discussions of local correlations of this type in the context of
nonperturbative, ``background-free" quantum gravity, it is a challenge to 
define suitable observables to convert statements like the above to more
quantitative ones, a line of enquiry we shall take up elsewhere.

\section{Discussion}\label{discuss}

The CDT prescription for constructing a theory of quantum gravity is 
extremely simple, namely, as the path integral over causal spacetime
geometries with a global time foliation. In order
to perform this summation explicitly, one introduces a grid of piecewise linear 
geometries, in much the same way as when defining the path integral in 
quantum mechanics. The action used is the Einstein-Hilbert action in the form
of the Regge action for piecewise linear geometries. Next, one rotates each of 
the Lorentzian geometries to Euclidean signature, and performs the 
path integral with the help of Monte Carlo simulations, thus restricting one 
to stay in the Euclidean sector. 
The key outcome is that in a certain range of bare coupling constants 
(``phase C") one observes a quantum universe which can be described as
an emergent four-dimensional cosmological ``background" geometry 
with superimposed quantum fluctuations, and a highly nonclassical
short-distance behaviour, as reflected in an anomalous spectral dimension,
$d_s \approx 2$, and some evidence of fractality. 

What is somewhat unusual compared to the standard lattice 
scenario is that the nontrivial infrared behaviour is observed for
a whole range of coupling constants. The purpose of this article was
to have a closer look at the geometric properties of the quantum
universe in this phase C, using various geodesic shell decompositions.
We were specifically interested in gathering further evidence for the
presence or otherwise of fractal features and in how these and some
global shape variables are influenced
by a change in the asymmetry parameter $\Delta$, which together with
the bare Newton's constant parametrizes the phase space of the 
underlying statistical model.
In previous work we have shown that for a specific choice of 
bare coupling constants in phase C (generic in the sense of not
being close to any phase transitions)
one can by a finite, global rescaling of the continuum cosmological proper 
time map the expectation value of the volume profile of the quantum
universe to that 
of a round four-sphere, that is, Euclidean de Sitter space. Does this 
picture change when we change the values of the bare coupling constants? 

In this paper, we have left the bare inverse gravitational constant unchanged
and varied the coupling constant $\Delta$ between the generic value 
$\Delta =0.6$ used previously and (almost) zero. 
We cannot go all the way to zero because there is a phase transition
just before we reach zero, and close to it our current 
Monte Carlo sampling becomes ineffective\footnote{A better sampling
which deals with the presence of 
vertices of very high order near the transition is currently under investigation.}. 
On the face of it, most of our results show a clear $\Delta$-dependence,
as illustrated by Figs.\ \ref{fig1}--\ref{fig5}. Only for small $\Delta$,
corresponding to $\alpha \approx 1$ are our results compatible with
a truly spherical universe. Is this in contradiction with earlier claims that
we observe a de Sitter universe throughout phase C? Not necessarily: 
as we have already alluded to in Sec.\ \ref{volshell}, it may be
that {\it continuum physics is invariant under a variation in} $\Delta$,
at least as long as we stay away from the phase transition.
 
\begin{figure}[t]
\centering
\includegraphics[width=0.75\textwidth]{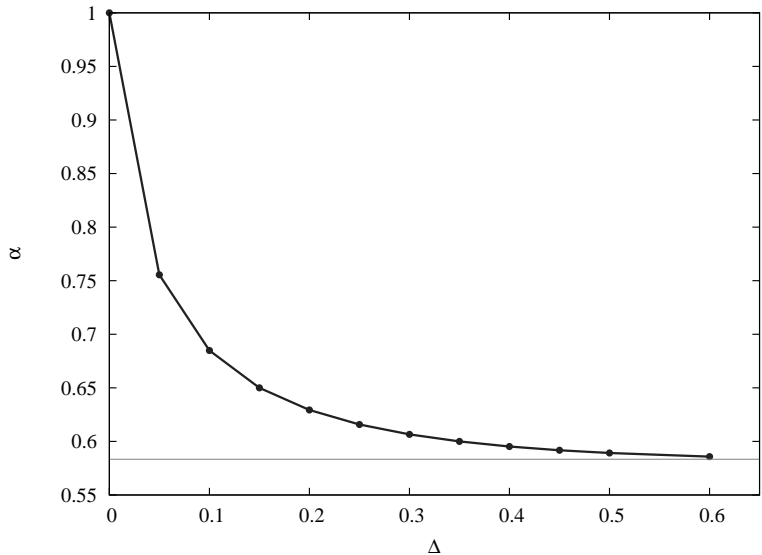}
\caption{\label{fig7} The asymmetry factor $\alpha$, 
plotted as a function of $\Del$, for $\kappa_0=2.2$.
The horizontal line is $\a=7/12$, the lowest kinematically allowed value 
of $\a$, where
the (3,2)-simplices degenerate because of a saturation of a triangle
inequality \cite{d4}.  
}
\end{figure}

Let us present some evidence that our current data is not in disagreement
with such a hypothesis. The key point is that 
$\Del$, which appears linearly in the action (\ref{actshort}) 
can be viewed as a choice of asymmetry
between space and time \cite{blp}. A more direct measure of this
asymmetry is given by the parameter $\alpha$, introduced originally in 
\cite{ori,d4} as the proportionality factor between the (squared)
length of time- and spacelike
edges, $a_t$ and $a_s$, according to $a_t^2 = \a a_s^2$. Now,
considering the relation between $\alpha$ and $\Delta$, plotted in
Fig.\ \ref{fig7}, one observes that a decrease in $\Delta$ is associated with an
increase in $\alpha$. In other words, a lattice unit in time direction corresponds
to an ever larger physical distance as $\Delta \rightarrow 0$. 
When taking this effect into account -- as one should -- when considering the
shape of the universe or the distribution $V_4(t,r)$, say, one sees that it could
potentially compensate for the differences for different $\Delta$ which
we observed when expressing our results in terms of fixed lattice units.
This would imply that the ``true" physics is unchanged under
variation of either $\alpha$ or $\Delta$ throughout phase C. 
The region where true sphericity is realized appears to be for $\alpha\approx 1$.
This also happens to be the region where the spatial diameter and the 
time extent, when measured in terms of discrete lattice units, are
approximately equal. 

It is difficult to convert this argument into a more quantitative 
statement about the shape of the spheroid, say,
because we do not at this stage have an independent way (other
than fitting the quantum universe to the round four-sphere) of establishing 
the relative
scaling between time and spatial distances in the continuum theory. 
The regularized theory is ambiguous when it comes to defining 
something like the ``timelike distance between spatial slices", not least
because of the singular nature of the piecewise flat geometries. 
The easiest definition is to take it to be unity in terms of discrete
lattice units, as is usually done. Alternatively, one could again take
the local, piecewise flat geometry literally and work out the true
geodesic distance of a point $x$ in the spatial slice at time $t+1$ to
the previous slice at time $t$ (which would lead one to
conclude that as a result of the specific geometry of the simplices and 
how they are glued together, 
this distance can vary between $c_1 a_s$ and $c_2 a_s$,
where $c_1,c_2$ are constants which themselves depend on $\a$).
Of course, this distance would also on average decrease when decreasing
$\alpha$, but would be distinct from the ``step distance" {\it at the
cut-off scale}. Nevertheless, 
one's expectation would be that different definitions of discrete distance
will give rise to equivalent notions of ``continuum distance", 
which differ at most by a global rescaling. However, it was exactly this
relative global scale we were trying to determine above. 

In summary, our hypothesis that continuum physics and geometry do
not change as the asymmetry parameter is changed continuously 
is not contradicted
by present measurements, but further corroboration will have to await
the study of finer-grained observables, which can distinguish spheroids
from true spheres. This also points to a potential
flaw in the way we have defined some of
our ``observables", like those depicted in Figs.\ \ref{fig1}--\ref{fig5}. 
In their definition, we simply treated time and space directions on an
equal footing (for example, when advancing shells in unit steps from
a given point). This can create a spurious $\Delta$-dependence.
For example, if our hypothesis is correct and the universe {\it is} a
round four-sphere, no matter where we are in phase C, one would
say that graphs like those for $\Delta = 0.25$ and $\Delta=0.6$ in 
Fig.\ \ref{fig2} cannot be counted as evidence for the presence of 
global anisotropy.

Once we cross the phase transition line and enter phase B, 
the situation changes  dramatically and there remains only a single
time slice which has a spatial three-volume
different from the minimal cut-off value -- four-dimensional
spacetime has completely disappeared! Presently it is unclear whether 
this happens
abruptly (corresponding to a first-order transition) or merely 
fast but smoothly (corresponding to a second-order transition). 
If the latter was the case, it would probably be inconsistent to maintain 
that the physical shape remained unchanged
all the way to the transition line. On the other hand, other scenarios 
may then suggest themselves, involving perhaps an asymmetric scaling 
of space and time along the lines envisaged in
Ho\v rava-Lifshitz gravity, see also \cite{to-app}.

Lastly, to return to the other one of our main themes, that of fractality,
our more detailed investigation finds little or no evidence of fractality 
when looking at a shell decomposition of spacetime. By contrast, 
when performing a shell decomposition {\it within} a given 
hypersurface (a shell or slice of constant $r$ or $t$), we have
confirmed earlier findings of a fractal structure \cite{blp} for
hypersurfaces of constant proper time and have
verified that they are also present for more general types of shells.   
We have gone one step further and found (at least qualitative) evidence
that the fractal structure of the hypersurface is propagated to a
neighbouring one, which means that it is not entirely an artifact confined
to a single, isolated shell. We do not yet understand the ramifications of
this result for the short-scale physics of the quantum universe. Most 
likely it is related to the anomalous spectral 
dimension observed in \cite{spectral} and obtained in both the 
asymptotic safety scenario \cite{spectral-as} and Ho\v rava-Lifshitz gravity
\cite{spectral-hl}. It would be interesting if this could be understood
in more detail.
 
\vspace{.5cm}
\noindent {\bf Acknowledgements.}
JA and RL are grateful to the Perimeter Institute for hospitality.
JJ acknowledges partial support through grant 182/N-QGG/2008/0 
``Quantum geometry and quantum gravity'', financed by the 
Polish Ministry of Science.
AG has been supported by the Polish Ministry of Science grants 
N~N202~034236 (2009-2010) and N~N202~229137 (2009-2012).
RL acknowledges support by the Netherlands
Organisation for Scientific Research (NWO) under their VICI
program.

\end{document}